\documentclass[useAMS,usenatbib]{mn2e}

\newcommand{\msun}{M$_{\sun}$}

\newcommand{\msuns}{M$_{\sun}~$}
\newcommand{\kmss}{km s$^{-1}~$}

\newcommand{\mnras}{MNRAS}
\newcommand{\apj}{ApJ}
\newcommand{\apjl}{ApJ}
\newcommand{\aj}{AJ}
\newcommand{\aap}{A\&A}
\newcommand{\araa}{ARA\&A}

\usepackage{graphicx}

\title[Limits on initial mass segregation in young clusters]{Limits on initial mass segregation in young clusters}
\author[N. Moeckel and I. A. Bonnell]{Nickolas Moeckel\thanks{E-mail:
nbm1@st-and.ac.uk} and Ian A.
Bonnell\\
SUPA, School of Physics and Astronomy, University of St Andrews, North Haugh, St Andrews, Fife, KY16 9SS}
\begin{document}

\date{Accepted 2009 March 20. Received 2009 March 18; in original form 2009 January 22}

\pagerange{\pageref{firstpage}--\pageref{lastpage}} \pubyear{2009}

\maketitle

\label{firstpage}

\begin{abstract}
Mass segregation is observed in many star clusters, including several that are less than a few Myr old.  Timescale arguments are frequently used to argue that these clusters must be displaying primordial segregation, because they are too young to be dynamically relaxed.  Looking at this argument from the other side, the youth of these clusters and the limited time available to mix spatially distinct populations of stars can provide constraints on the amount of initial segregation that is consistent with current observations.  We present {\it n}-body experiments testing this idea, and discuss the implications of our results for theories of star formation.  For system ages less than a few crossing times, we show that star formation scenarios predicting general primordial mass segregation are inconsistent with observed segregation levels. 
\end{abstract}


\section{Introduction}
Many star clusters show some degree of radial mass segregation, with massive stars over-represented in the inner region of the cluster.  This is expected in bound clusters that are older than a dynamical relaxation time, such as the Arches near the galactic center, which shows an almost-flat mass function near the center \citep{kim06}.  Theoretical explanations and numerical explorations of this segregation process are numerous \citep[e.g][]{spitzer69, spitzer75, khalisi07, portegies-zwart07}.  

Younger clusters, with ages less than a few Myr, are not expected to display the same degree of segregation, as they are only a few crossing times old and should not have had time to segregate dynamically.  Despite this, there are several examples of young (or even embedded) clusters exhibiting mass segregation.  Examples with ages less than $\sim 5$ Myr include: Mon R2 \citep{carpenter97}; IC 1805 \citep{sagar88}; NGC 1893 \citep{sharma07}; NGC 6530 \citep{mcnamara86}; NGC 6231 \citep{raboud98a}; and the Orion Nebula Cluster (ONC) \citep{hillenbrand98}.  Some of these arguably show a general trend of segregation, with more massive stars progressively more centrally located.  Others, such as Mon R2, show little evidence or this type of segregation, yet still have a central concentration of the most massive stars.  The youth of these clusters relative to their relaxation time makes the process of dynamical segregation less likely, and this timescale argument has been used as evidence that primordial segregation has played a role \citep{hillenbrand98,bonnell98,raboud98a}.

Most numerical explorations of dynamical mass segregation have focussed on long timescales \citep[10s--100s of crossing times, e.g.][]{khalisi07}.  \citet{bonnell98} explored the short-timescale {\it n}-body evolution of a variety of initial cluster setups, concluding that the formation of a Trapeziumesque system in the center of an ONC-scale cluster is unlikely to be realized by two-body relaxation alone, and that the most massive stars in Orion would have to have formed within the inner $\sim 25$\% of the cluster to find themselves in the center at the present day.  \citet{mcmillan07} performed {\it n}-body experiments of merging sub-clusters, showing that the merger products can exhibit a more advanced state of segregation than their age would suggest, somewhat ameliorating the timescale problem as the memory of rapid dynamical evolution in small-{\it n} clusters is apparently retained when they merge.  Nonetheless, it remains unclear how important primordial segregation is relative to later dynamical evolution.  

While mass segregation is a robust and well studied result of dynamical interactions in a cluster, the usual starting state for these explorations is an initially unsegregated cluster.  It is unclear how appropriate such initial conditions are for very young clusters.  Theories of massive star formation \citep[e.g.][]{bonnell01a,mckee03} can easily be interpreted to imply that the mass of a star depends on its location in such a way that the structure of a cluster when gravitational dynamics become dominant is highly segregated.  In this case, mass segregation is not the eventual result of a cluster trying to achieve equipartition, but the (perhaps unstable) starting point.  
In this paper we approach mass segregation from the other side, exploring the evolution of very young clusters from a primordially segregated state.  We present {\it n}-body experiments covering a range of initial cluster setups and compare the distributions of stars to observations of clusters only a few crossing times old, and we discuss the implications of our results for theories of star formation.

\section{Numerical details}

To explore this problem, we performed a suite of {\it N}-body experiments using the code {\sc NBODY6} \citep{aarseth03}.  The code uses a fourth-order Hermite integrator to advance the self-gravitating system.  The energy error $\Delta E / E_0$ over the course of each simulation was held under $10^{-4}$.

The stars in the simulations have masses in the range 0.08 -- 50.0 \msuns drawn from the IMF given by \citet{kroupa01}:
\begin{equation}
\label{imf}
  \xi(m) \propto \left\{
  \begin{array}{cc} 
    m^{-1.3},  &0.08 \le m/M_{\sun} < 0.5,~\\
    m^{-2.3},  &0.5 \le m/M_{\sun} \le 50.0 \\
  \end{array}
  \right.
\end{equation}

Throughout, we neglect the effects of an initial binary population or the remnants of the cluster's gas reservoir.  In order to begin with a mass-segregated cluster, we set up the positions of the stars in a mass-dependent fashion.  Each star is assigned a uniformly distributed location within a sphere of radius
\begin{equation}
  r_{max}(M) = \left( \frac{M}{M_{min}} \right)^{-q},
  \label{radiusequation}
\end{equation}
where $M$ is the mass of the star being initialised and $M_{min}$ is the minimum mass in the cluster.  the index $q$ controls the level of segregation; with $q=0$ all stars have the same maximum radius independent of mass, while $q=1$ yields a very strict radial ordering by preventing massive stars from being located in the cluster outskirts.  At this point the stars are spatially arranged with a radial mass (or number) density that does not correspond to any standard scenario.  Working from $r = 0$ outward, each star in turn has its radius adjusted to correspond to some underlying distribution.  The first step thus effectively ranks the stars radially as a function of their mass, while the second step adjusts the initial conditions to a physically motivated radial profile.

Sorting the stars radially by mass has a drastic effect on the dynamical structure of the cluster.  The mass profile of a heavily mass segregated cluster with an isothermal number density profile is much more centrally condensed than an unsegregated cluster.  To encompass a wide range of initial parameters, we perform two sets of simulations.  In the first set, we adjust the radii of the bodies so that the {\em number} density is uniform with radius, $n(r) = n_{const}$.  In the second set, we adjust the radii so that the {\em mass} density is an isothermal sphere, $\rho(r) \propto r^{-2}$.  This segregation scheme preserves the overall cluster IMF.  We note that \citet{subr08} recently devised a method for generating stable, virialized clusters with a parameterized degree of mass segregation, and we defer consideration of this method to the discussion section.

In each set, we performed three series of 100 experiments, each with a different level of initial mass segregation:
\begin{description}
  \item {\em Series A -- no initial segregation}: we set $q=0$ in equation \ref{radiusequation}, so that the radial distribution is independent of mass.
  \item {\em Series B -- moderate initial segregation}: we set $q=0.15$ in equation \ref{radiusequation}.  At this value, stars more massive than $\sim 5$ \msuns are restricted to lie inside half the cluster radius.  After adjustment to an underlying distribution, this changes somewhat.
    \item {\em Series C -- heavy initial segregation}: we set $q=1$ in equation \ref{radiusequation}.  This value of $n$ effectively sorts the stars almost strictly by mass, with the most massive stars in the cluster center.
 \end{description}

Each cluster begins with 1000 stars, and we run each cluster for 6 crossing times.  We scale all of the clusters so that the initial half-mass radius $r_{hm} = 0.5$ pc, the rms velocity in the clusters is $\sigma \approx 1.5$ \kmss for each series, and the initial crossing time $\tau_c = 2 r_{hm}/\sigma \approx 0.65$ Myr.  With this scaling we are simulating the clusters for more than 3.5 Myr, encompassing a more-than-sufficient time to compare to young clusters such as the ONC.

\section{Uniform Number Density Results}
\begin{figure}
 \includegraphics[width=84mm]{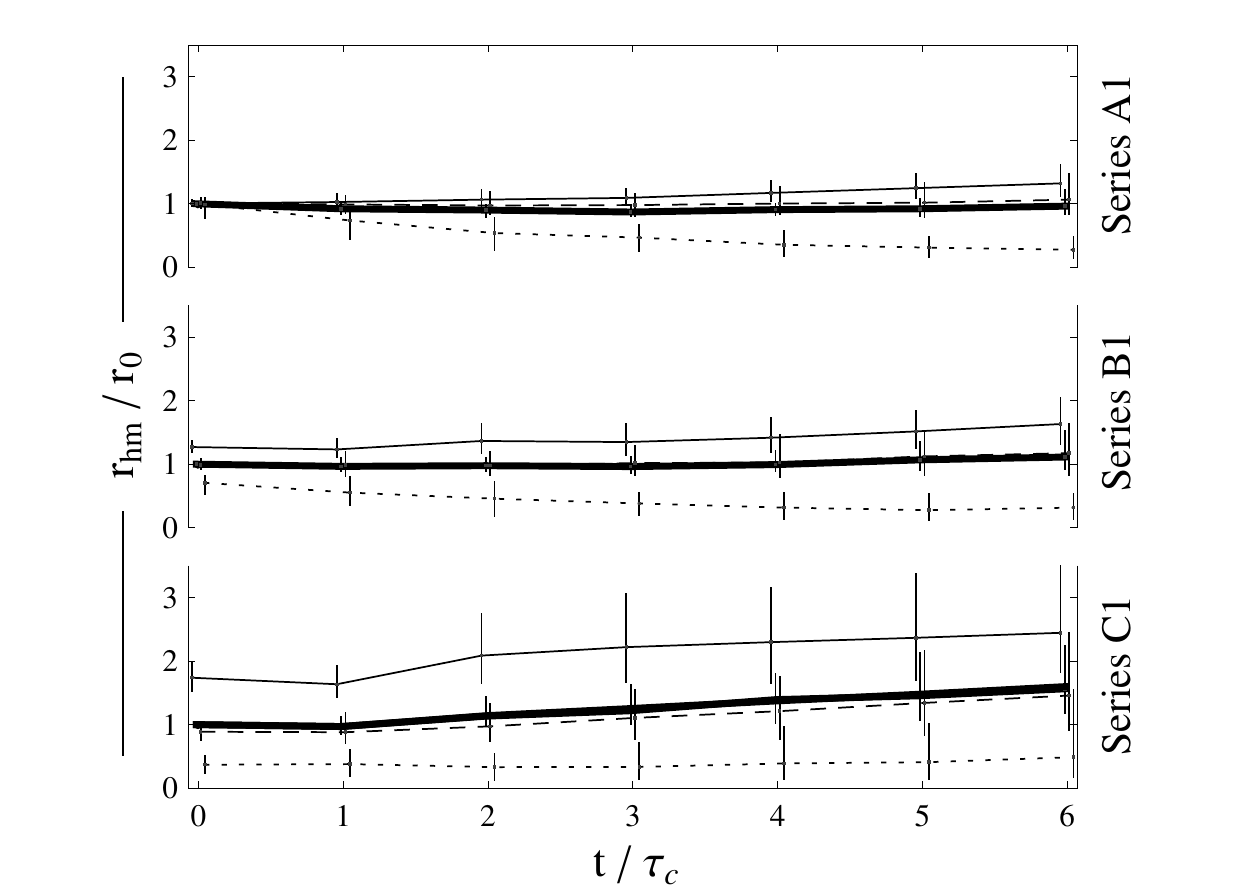}
 \caption{Evolution of the half-mass radius $r_{hm}$ for each of the uniform number density series, relative to the initial half-mass radius $r_0$.  The median value of all the simulations is plotted, with error bars showing the upper and lower 95th percentile values.  Horizontal offsets separate the error bars.  The different lines correspond to the following subsets of the system: {\em thick}: all stars; {\em solid}: $0.08 \le m/M_{\sun} < 1.0$;   {\em dashed}: $1.0 \le m/M_{\sun} < 5.0$;   {\em dotted}-- $5.0 \le m/M_{\sun} \le 50.0$.
 \label{RhmUniFig}}
\end{figure}  

This set of simulations began from an initially uniform number density, upon which mass segregation was imposed with no concern for what this does to the mass distribution.  All of the simulations in this set begin in virial equilibrium.  In our analysis of the simulations, we split the stars into three groups: low-mass stars,  $0.08 \le M_{\star}/M_{\sun} < 1.0$; moderate-mass stars, $1.0 \le M_{\star}/M_{\sun} < 5.0$; and high-mass stars, $5.0 \le M_{\star}/M_{\sun} \le 50.0$.  

\subsection{Radial Evolution}
We begin by considering the overall evolution of the clusters in this set, measured by the half-mass radii of the different mass components.  In Figure \ref{RhmUniFig} we show the evolution of the half-mass radius for each series, broken into mass bins and over the whole cluster.  Plotted is the median value over all the simulations in a series.  Examining the trends in these quantities for series A1 (initially unsegregated), it is apparent that after only a few crossing times some degree of mass segregation is setting in.  While the overall half-mass radius remains fairly constant, massive stars have moved inwards so that after 6 crossing times their $r_{hm}$ is $\sim$30\% of its initial value.  This evolution occurs as the least-massive stars, and to a lesser extent the moderate-mass stars, are driven outward.  This degree of rapid segregation is consistent with \citet{bonnell98} and \citet{subr08}.

Comparing the evolution of the half-mass radii for series B1 (moderate initial segregation) with series A1, the trends are similar.  With the exception of the high-mass bin, the changes in $r_{hm}$ for each component in series B1 closely match the unsegregated case but offset by the initial segregation.  The inward evolution of $r_{hm}$ for massive stars appears to saturate at $\sim 30$\% of the total cluster value, so that after $t = 5~\tau_c$ series A1 and B1 have similar values.  The overall half-mass radius is nearly constant over the course of the simulations.

The overall cluster evolution of series C1 (heavy initial segregation) is qualitatively different than the other two uniform number density experiments, as the cluster expands from its initial configuration.  Rather than the massive stars moving inwards as in series A1 and B1, the massive population remains radially stationary.  The low-mass stars adjust their location quickly outwards between 1 and 2 crossing times, and the moderate-mass stars follow a general outward trend with time.

\subsection{Cumulative Distributions}
\begin{figure}
 \includegraphics[width=84mm]{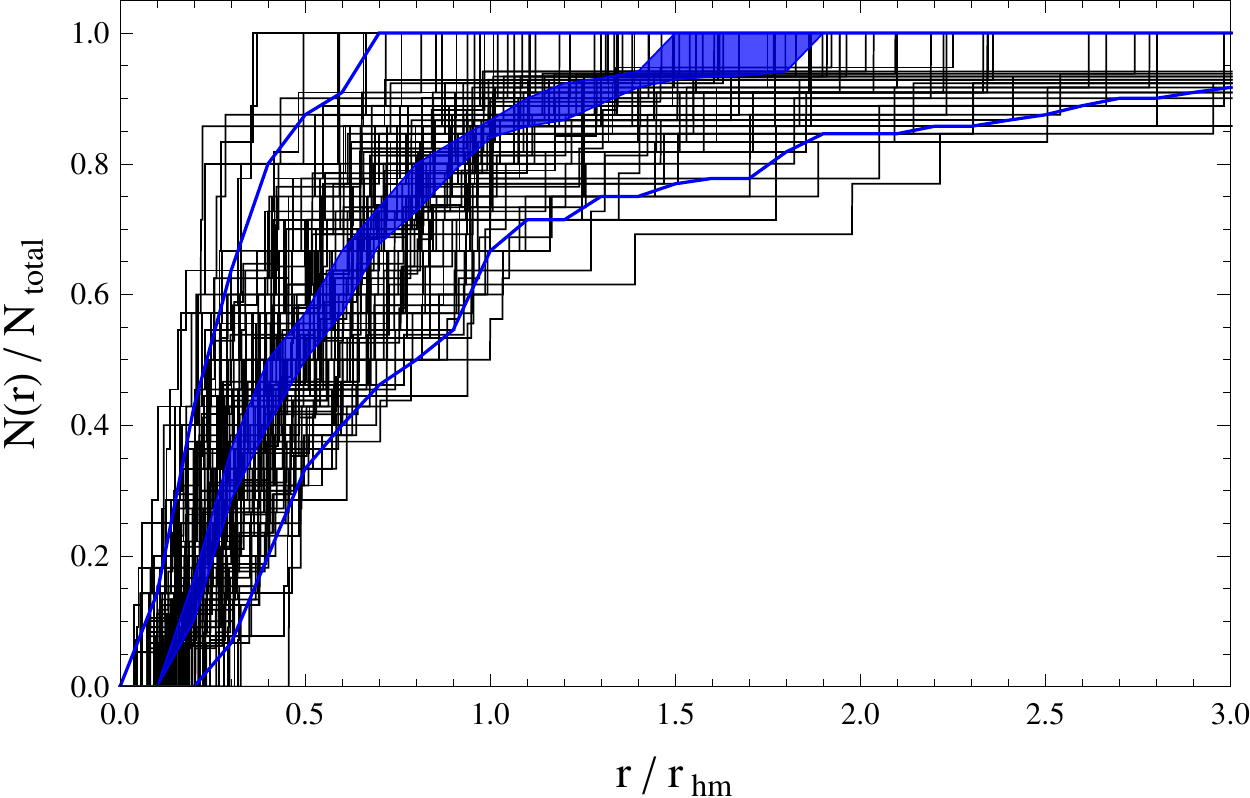}
 \caption{Demonstration of our reduction scheme, for stars more massive than 5 \msuns in series A at $t = 6 \tau_c$.  The cumulative number distribution with radius for each of 100 runs are shown as black lines.  The dark shaded region shows the 95\% confidence interval of the location of the underlying distribution's median, found via a bootstrap resampling of the simulation data.  The outer lines mark the uper and lower 95th percentiles of the simulated distribution, to give an idea of the scatter in the ensemble.  Further figures in this work will show only these lines and the median.\label{MedExpFig}}
\end{figure}

For each mass range, we consider the cumulative number distribution of stars.  Comparing these distributions gives a measure of the mass segregation in the systems.  We treat the results of each series of initial conditions as an ensemble, and attempt to recover the underlying distribution for each mass bin at different times.

In Figure \ref{MedExpFig} we illustrate this process.  In this example, from series A at t = 6 $\tau_c$, we plot for all 100 experiments the cumulative fraction of high-mass stars as solid lines.  This population of stars displays the largest spread in the the final distribution within one series, due to the relatively low-number statistics (the mean number of stars in this mass range is 11, with standard deviation $\sim 3$).   The spread of the cumulative distribution function is asymmetrical, so we use the median as the characteristic value for the ensemble of simulations.   The dark shaded region shows the 95\% confidence interval of the {\em underlying distribution's} median, obtained by a bootstrap resampling of the simulation data.  The thick lines mark the 95th percentiles of the simulated ensemble, and give an indication of the spread in simulated values.  In the rest of this discussion we plot only these quantities, rather than the results of each individual run.

\begin{figure*}
 \includegraphics[width=160mm]{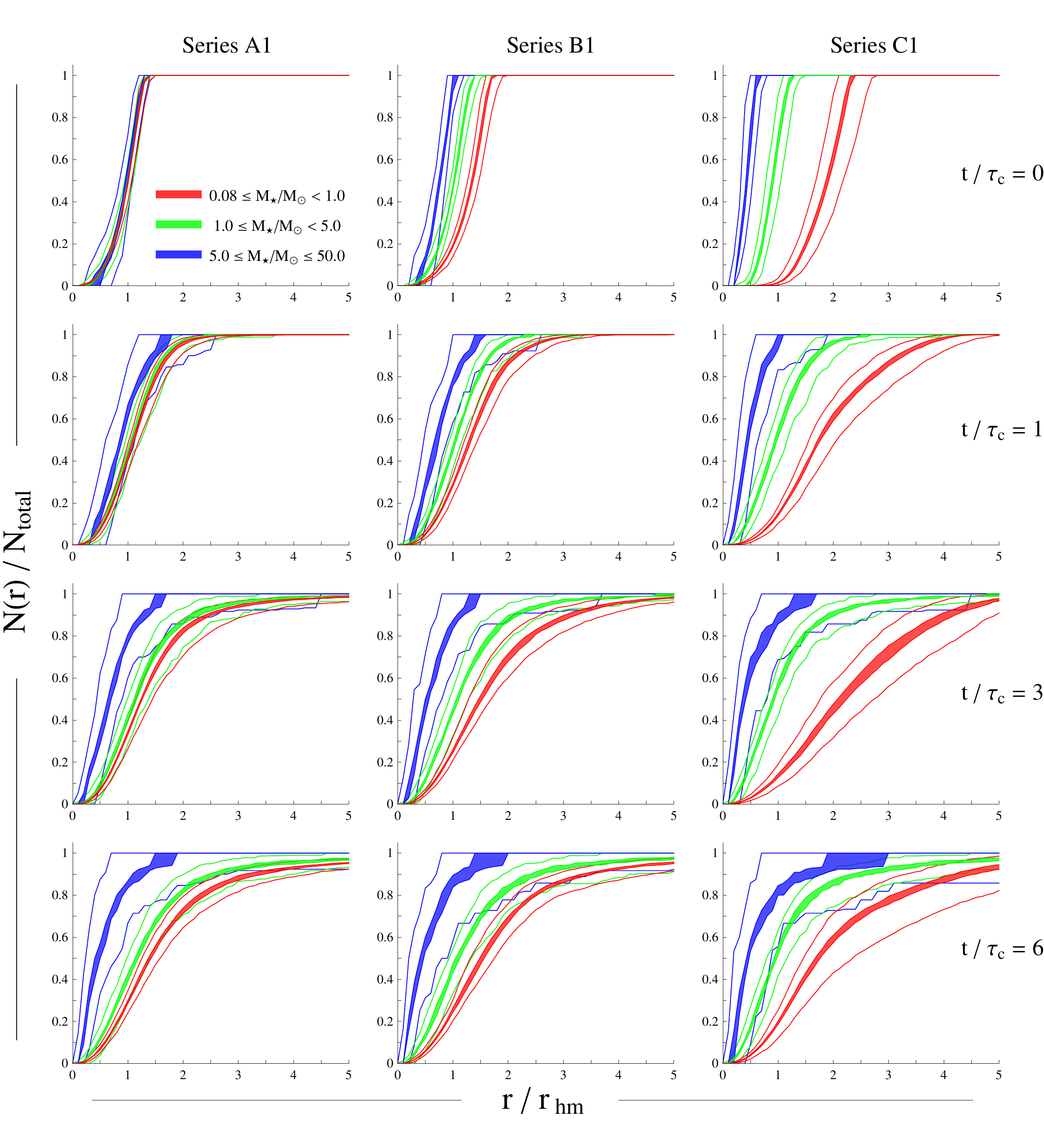}
 \caption{Evolution of the cumulative number distribution of three mass components for the series with uniform number distributions.  Columns are labeled with the series they represent, and rows are labeled with time increasing downwards.  {\it Filled region}:  95\% confidence limit on the median values of the distribution; {\it lines}: 95th percentile limits of the distribution.
 \label{UniformResultsFig}
 }
\end{figure*}

\subsubsection{Series A1--Uniform number density, no segregation}
Figure \ref{UniformResultsFig} shows the time evolution of the distributions of the three mass components for all three series in the uniform set, with the left column showing series A1.  In the four panels we show the cumulative distribution of each mass component at times $t = 0, 1, 3$ and 6 $\tau_c$.  The radius in each plot is shown as a fraction of the half-mass radius {\em at that time}, rather than at the initial time.  We make this choice to show what an observer might see looking at a cluster.  With knowledge only of the current state, measuring radius relative to an unknown and irrelevant initial mass distribution would be impossible.

At the beginning of the cluster evolution, the stars in each mass bin are uniformly distributed throughout the cluster, evidenced by the identical form of their initial cumulative distribution functions.  After 1 crossing time the 
shape of the cluster has changed somewhat, and even at this early time there is a hint of mass segregation of the high-mass stars.  The distributions of the different components are mostly still overlapping, however.  By 3 crossing times the segregation of the massive stars has become more pronounced, while the moderate- and low-mass stars show a small separation in their distribution but remain quite similar.  Between 3 and 6 crossing times the separation between the low- and moderate-mass distributions becomes clearer, though their medians are nearly coincident with the other's lower and upper 95th percentile respectively.  From $t = 3 \tau_c$ onward, the high-mass distribution is completely separated with a high degree of confidence from the other two components in the inner regions of the cluster, and the median value is separated  from the extremes of their distributions at all radii.

\subsubsection{Series B1--Uniform number density, moderate segregation}
The middle column of Figure \ref{UniformResultsFig} shows the cumulative distribution plots for series B1.  In this model the massive stars are all initially placed within the half-mass radius, but there is still a radial overlap among radial distributions of the three mass components (i.e. low-mass stars are present in the innermost region of the cluster; this is in contrast to series C1).  Because of the central concentration of the massive stars, the half-mass radius is a smaller fraction of the total cluster radius.  The evolution is very similar to that seen in series A1, with the separation between the low- and high-mass distributions getting clearer with increasing time, but with more advanced levels of segregation throughout.  The main difference between this series and the unsegregated case is the separation between the low- and moderate-mass stars.  The initial difference in these distributions is not erased over the course of the simulations.  This is best seen by comparing the results from series A1 and B1 at $t=3\tau_c$.  In the unsegregated runs, the medians of the low- and moderate-mass distributions are marginally separate, but the lower 95th percentile of each distribution is  nearly the same.  In contrast, the lower-mass components are still clearly separated at this time in series B1, with the respective upper and lower 95th percentiles nearly coincident for much of the radial range.  The difference between these runs is less distinct after 6 crossing times.

\subsubsection{Series C1--Uniform number density, heavy segregation}
The cumulative distributions for this series are shown in the right-hand column of Figure \ref{UniformResultsFig}.  The extreme nature of the initial segregation is apparent in the lack of radial overlap among the three populations; the moderate-mass stars only begin to appear outside of the high-mass stars, and the low-mass stars are almost exclusively located outside the half-mass radius.  Up through $t=3\tau_c$ the evolution of the distributions is mostly dominated by the outward movement of the low-mass stars, with the separation between the medians of the different distributions remaining clear at all radii, and the extremes of the distributions distinguished inwards of $r_{hm}$.  By $t=6\tau_c$ the moderate- and high-mass distributions have moved closer together, though inwards of $\sim 2 r_{hm}$ they remain clearly separated.

The point of this series is to test the efficiency of mixing from such an extreme setup.  If this case showed thorough mixing after only a few crossing times, than it would seem that observations of mass segregation in young clusters hold little value in telling us where stars of different masses form.  However, the clear signal of segregation is retained through at least 6 crossing times, corresponding to several Myr for a parsec-scale cluster.

\section{Isothermal Mass Density Results}
In these experiments the radii of the stars are adjusted so that the underlying mass density is an isothermal sphere, $\rho(r) \propto r^{-2}$.  Thus the mass profiles and total radial extent of each of the three series are identical, with the outer cluster radius at $2 r_{hm}$, though the number densities are much different.  The non-segregated case has a number density that is also isothermal, while the segregated cases have shallower number density profiles that are radially-dependent .  This set of experiments can be thought of as representative of clusters forming from the same progenitor molecular cloud, with different star formation mechanisms.  The results presented all began in a state of virial equilibrium, though we also ran these as slightly sub-virial systems with a virial parameter $q = E_{kin} / |E_{pot}| = 0.4$.  Differences are noted where appropriate below.

\subsection{Radial Evolution}
\begin{figure}
 \includegraphics[width=84mm]{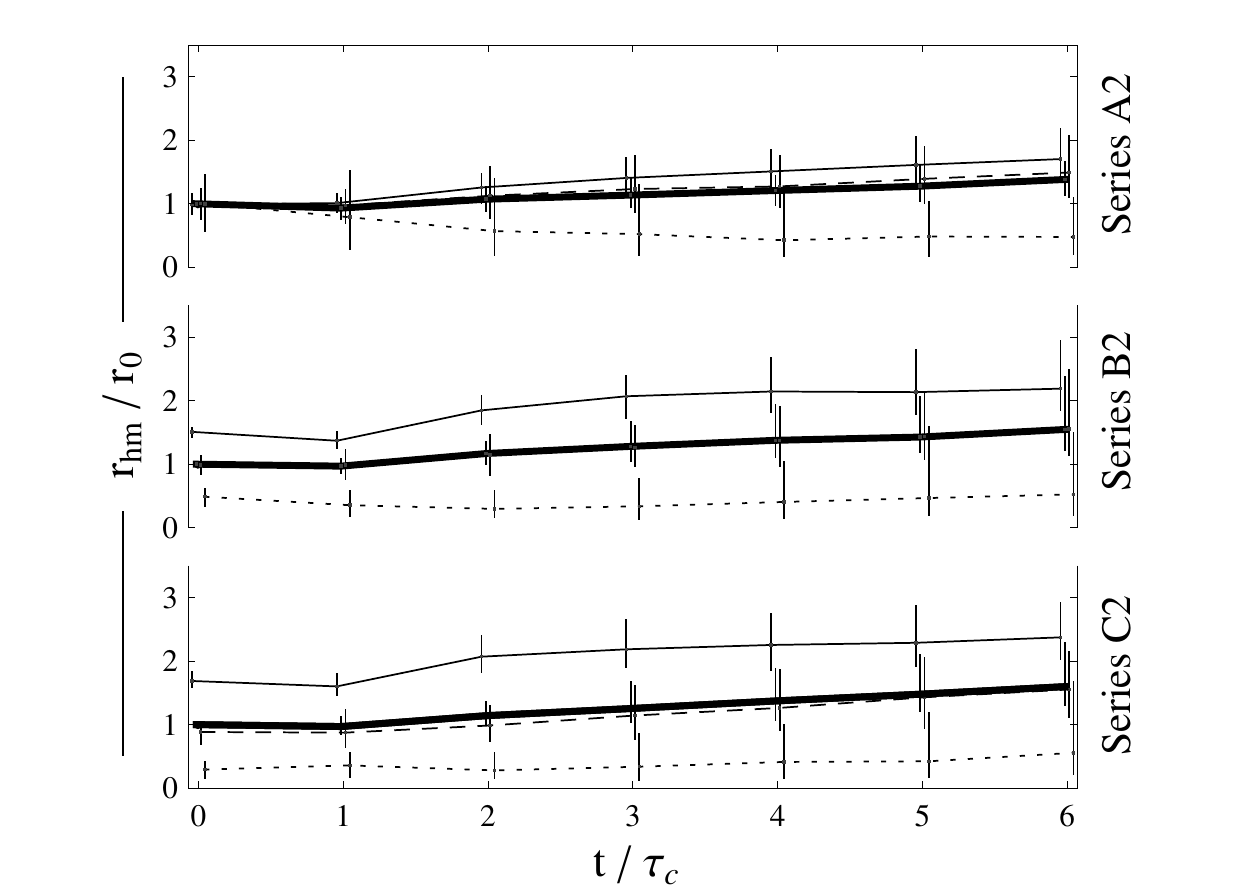}
 \caption{Evolution of the half-mass radius $r_{hm}$ for each of the isothermal mass density series, relative to the initial half-mass radius $r_0$.  The median value of all the simulations is plotted, with error bars showing the upper and lower 95th percentile values.  Horizontal offsets separate the error bars.   The different lines correspond to the following subsets of the system: {\em thick}-- all stars; {\em solid}: $0.08 \le m/M_{\sun} < 1.0$;   {\em dashed}: $1.0 \le m/M_{\sun} < 5.0$;   {\em dotted}: $5.0 \le m/M_{\sun} \le 50.0$.
 \label{RhmIsoFig}}
\end{figure}

We again begin by examining the half-mass radii for this set, shown in Figure \ref{RhmIsoFig}.  The top panel shows series A2.  This series is quite similar to the uniform number density case, though with a small degree of overall cluster expansion over 6 crossing times.  The same trend of inward migration for massive stars with outward movement of the low mass stars is seen, and to a similar extent as in series A1.

The main difference between the half-mass radii evolution in series B2 versus B1 is the early outward movement of the low mass stars between 1 and 2 $\tau_c$.  The root cause of this, which is also seen in series C2 and was present to some extent in series C1, is the assumption of a mass-independent, isothermal velocity distribution. When mass is concentrated to the center of the cluster, the magnitude of the system's potential energy increases relative to an unsegregated cluster, and the velocity dispersion consequently increases.  The tendency for fast-moving stars at the cluster outskirts to drift away from the cluster is exacerbated somewhat by not imposing a radial dependence on the velocity dispersion, as in a Plummer or King profile.  Though this would bolster the stability of the radial mass profile, observations tend to favor single-dispersion descriptions with little evidence for radial dependence in young clusters \citep[for instance in the ONC,][]{hillenbrand98}.

In series C2 (an as in series B2 and C1) there is a rapid outward adjustment of the low-mass stars between 1 and 2 $\tau_c$, after which the evolution of this series is very similar to the uniform number density case.  A moderate increase of the overall cluster radius takes place as the massive star distribution remains largely stationary.

\subsection{Cumulative Distributions}
\begin{figure*}
 \includegraphics[width=160mm]{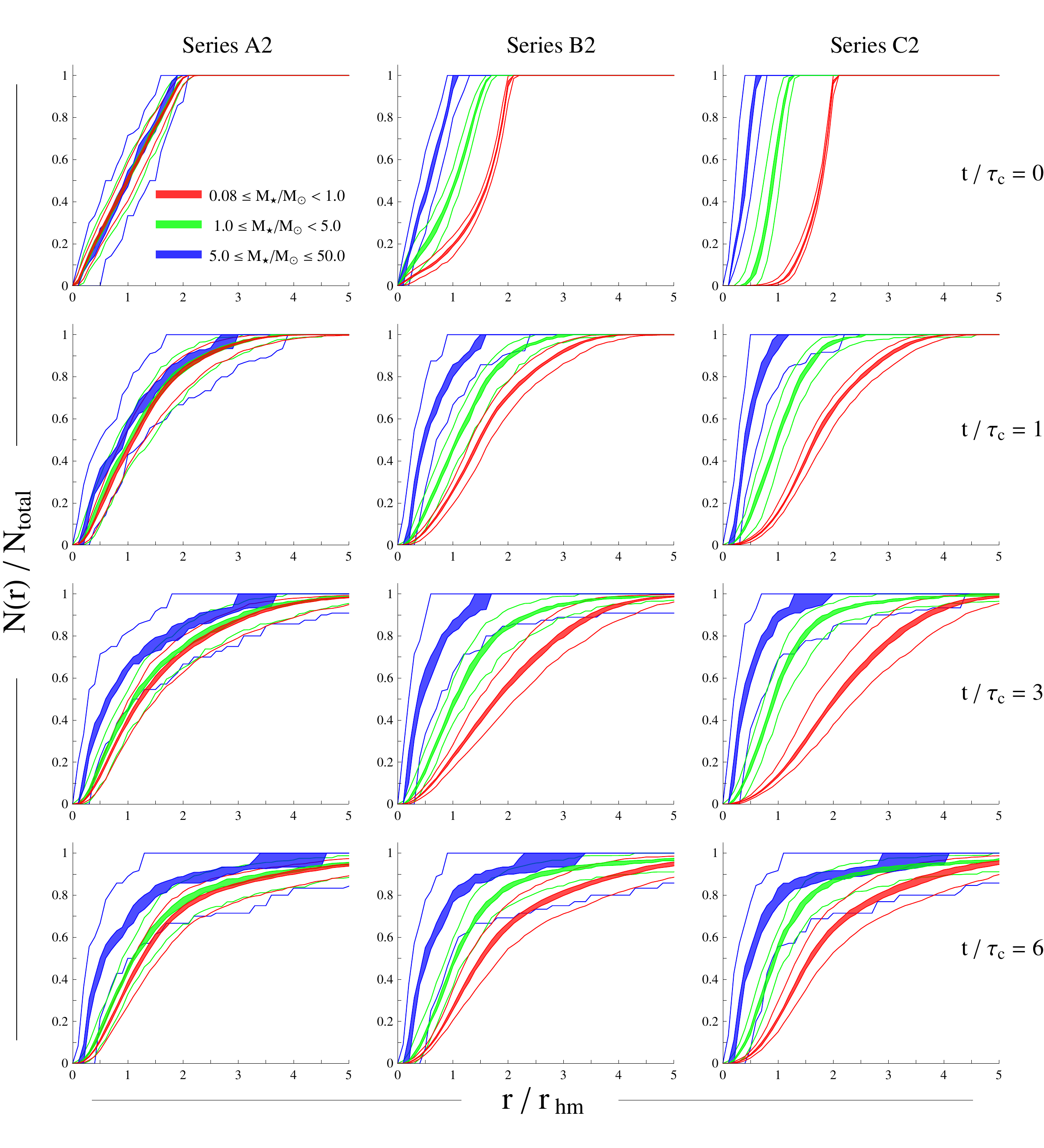}
 \caption{Evolution of the cumulative number distribution of three mass components for the series with isothermal mass distributions.  Columns are labeled with the series they represent, and rows are labeled with time increasing downwards.  {\it Filled region}:  95\% confidence limit on the median values of the distribution; {\it lines}: upper and lower 95th percentiles  of the distribution.
 \label{IsothermalResultsFig}}
\end{figure*}

In Figure \ref{IsothermalResultsFig} we show the evolution of the mass distribution for this set in the same fashion as for the uniform number density results.  
\subsubsection{Series A2--Isothermal mass density, no segregation}
The isothermal mass distribution for this set is reflected in the linear growth of the cumulative fraction for each mass component.  After 1 crossing time there is no real separation between the components.  By 3 crossing times the high-mass distribution is beginning to separate from the lower mass stars, which continues through $t = 6 \tau_c$.  As in the previous setup with no segregation, only marginal separation between the distributions of the low- and moderate-mass stars is seen.

\subsubsection{Series B2--Isothermal mass density, moderate segregation}
The middle column of Figure \ref{IsothermalResultsFig} shows the cumulative distribution evolution for this series.  As in the uniform number density case, the massive stars are initially all within the half-mass radius, but there is still radial overlap among the components.  The initial separation between the components never gets erased through $6 \tau_c$, though by the end of the runs the extremes of the distributions are overlapping.  The medians, however, remain clearly separated at all radii through 3 crossing times, and within $2 r_{hm}$ at 6 crossing times.

When evolved from slightly sub-virial initial conditions, the evolution of this series is only mildly affected.  The period between $t = 0$ and $t = 1 \tau_c$ is marked by overall contraction of the cluster as it seeks equilibrium.  This contraction is seen in all three mass components.  From that point onwards, the behavior of the sub-virial system is as described for the virial case, with expansion of the low-mass component and slow growth of the overall half-mass radius.  The changes to the cumulative distribution plots are minimal.

\subsubsection{Series C2--Isothermal mass density, heavy segregation}
Like the other segregation levels, the evolution of this series shows a close similarity to its uniform number density counterpart.  The initial conditions again show an almost-total amount of segregation, with the moderate-mass stars only appearing outside of the massive stars, and the low-mass stars excluded from inside the half-mass radius.  The separation between the moderate- and high-mass distributions grows less distinct with time and increasing radius, but they remain separate at a level comparable to series B2.  The low-mass distribution is separate from the higher mass stars even at the 95th percentile level out to $\sim 2 r_{hm}$ at all times.  

As with the moderate segregation case, running this system sub-virially in an effort to stem the expansion of low-mass stars has a minimal impact on the overall evolution, with the only real difference during the first crossing time as the cluster contracts to $\sim 70$\% of its initial size before evolving as the virial case does.  The cumulative distribution plots are likewise almost unaffected.  
\begin{figure}
 \includegraphics[width=84mm]{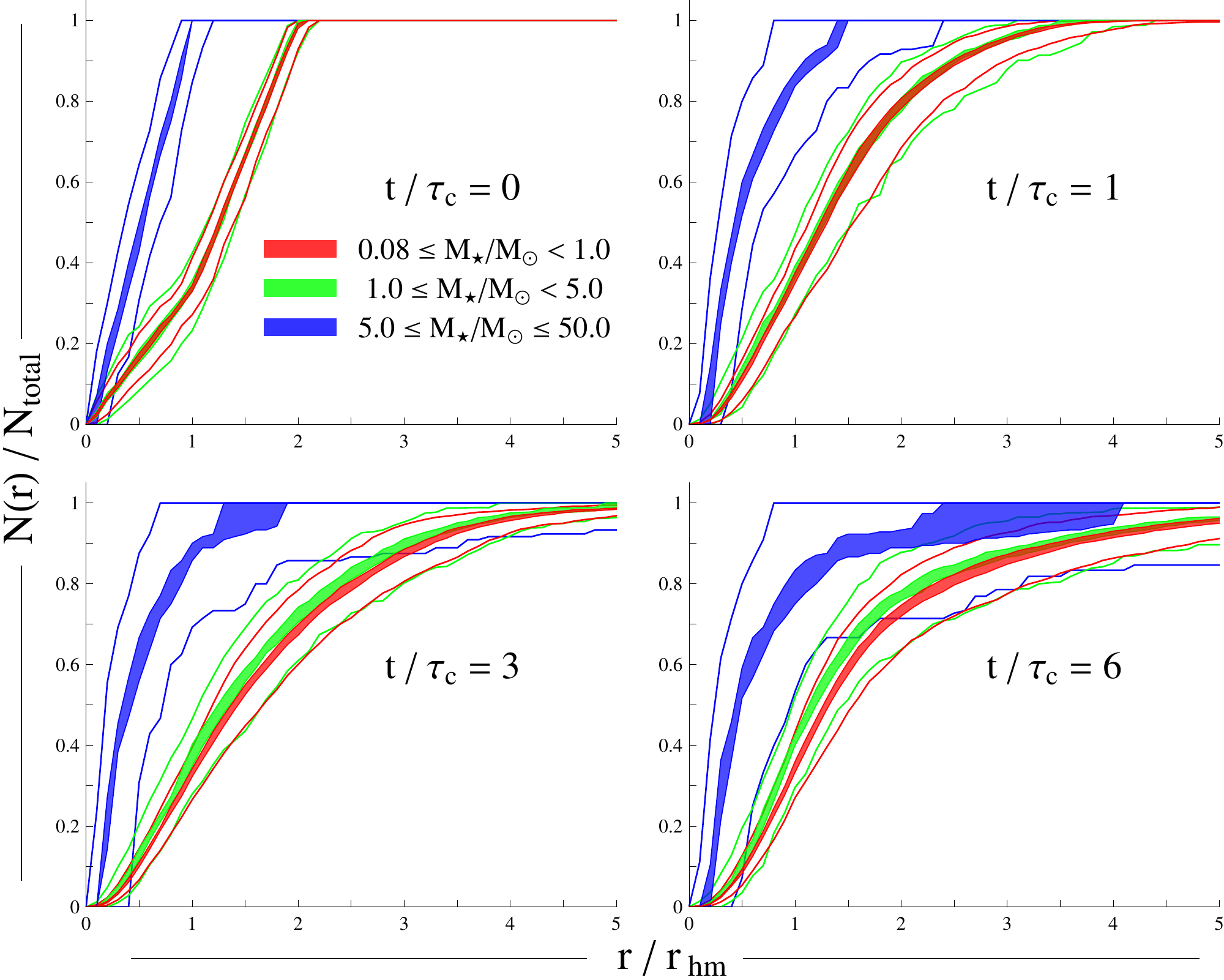}
 \caption{Evolution of the cumulative number distribution of three mass components for series D, with an isothermal mass distribution and initial segregation of only the massive stars.  Times are labeled on the figure.  {\it Filled region}:  95\% confidence limit on the median values of the distribution; {\it lines}: upper and lower 95th percentiles  of the distribution.
 \label{DResults}}
\end{figure}

\subsubsection{Series D--Isothermal mass density, only massive stars segregated}
If massive stars form via a fundamentally different process than low-mass stars (for instance, if stars earlier than type B require different conditions to accrete past the obstacle of their own radiation), it could be the case that only these stars would be initially segregated.  This idea is supported by dynamical studies \citep{bonnell98}, perhaps some observations \citep[e.g. the Trapezium][]{hillenbrand98}, and is implicit in some theories of massive star formation \citep[e.g. the fiducial model of][]{mckee03}.
  When comparing these results to observations, it will therefore prove useful to consider a setup in which only the massive stars are initially segregated.  We use the isothermal mass density scheme to test this scenario, and the results are shown in Figure \ref{DResults}.  All of the massive stars are initially inside the half-mass radius, while the distributions of low- and moderate-mass stars are identical and extend throughout the cluster.  Throughout the run the massive stars remain highly segregated.  The low- and moderate-mass stars segregate at a level comparable to the unsegregated cases A1 and A2.  A scenario in which the most massive stars form in a compact configuration at a cluster center, as the Trapezium appears now and as W3 IRS 5 appears at a younger age \citep{megeath05}, is difficult to distinguish from initially unsegregated initial conditions if one looks only at the relative distributions of the low- and moderate-mass stars.

\section{Discussion of Numerical Results}
\begin{figure}
 \includegraphics[width=84mm]{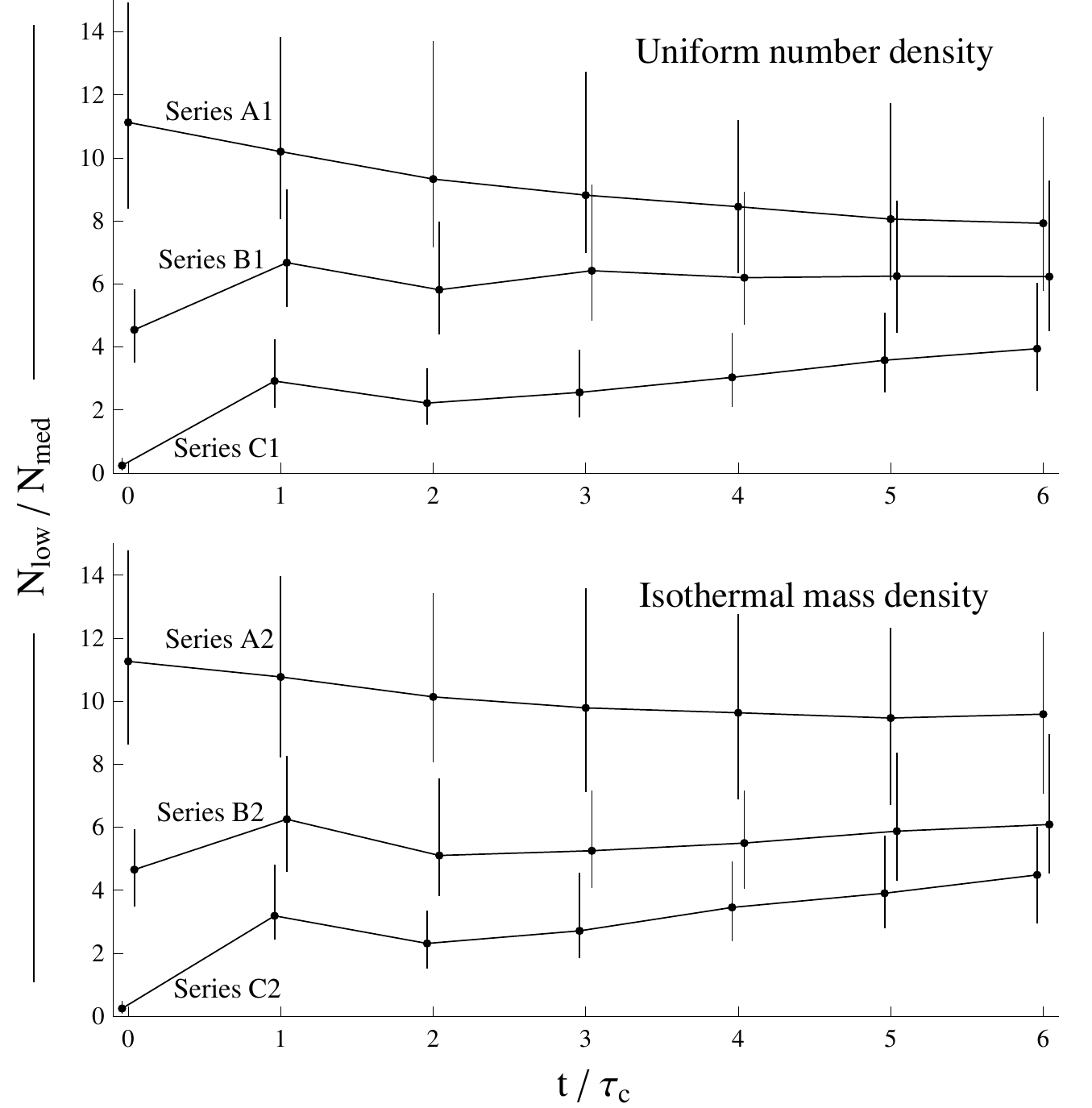}
 \caption{The ratio of the number of low-mass to moderate-mass stars within the half-mass radius, $N_{low}/N_{med}$.  The median value for each series is plotted, with error bars showing the upper and lower 95th percentile values.  The small horizontal offsets serve to separate the error bars, all times are measured at integer crossing times.
 \label{LowMedRatioFig}}
\end{figure}
Examining the distributions in Figures \ref{UniformResultsFig} and \ref{IsothermalResultsFig}, the clearest discriminant between the different levels of segregation appears to be the distribution of the low- and moderate-mass stars.  The unsegregated cases show only a very marginal separation between these distributions, while the runs with initial segregation maintain the initial distinction throughout the simulations.  To put a number to the visual difference between these populations, we use the ratio of low-mass to moderate-mass stars, $N_{low} / N_{med}$, for all stars located inside the half-mass radius.  While it could be tempting to compare these results to observed clusters, this ratio is extremely sensitive to survey completeness, and the cumulative distribution is probably a preferable means of comparison to reality, which we examine further below. 

Shown in Figure \ref{LowMedRatioFig} is the median value of this ratio for each series, with the error bars showing the upper and lower 95th percentiles.  While there is a fairly large spread in this quantity, the median values remain distinct at all times.  For the uniform number density set, the moderate-segregation case (series B1) begins to lose its distinction from the unsegregated case (series A1), also noted in the discussion of the cumulative distribution.  The isothermal mass density series retain their difference for longer times, also evident in the cumulative distribution plots.

\begin{figure*}
 \includegraphics[width=140mm]{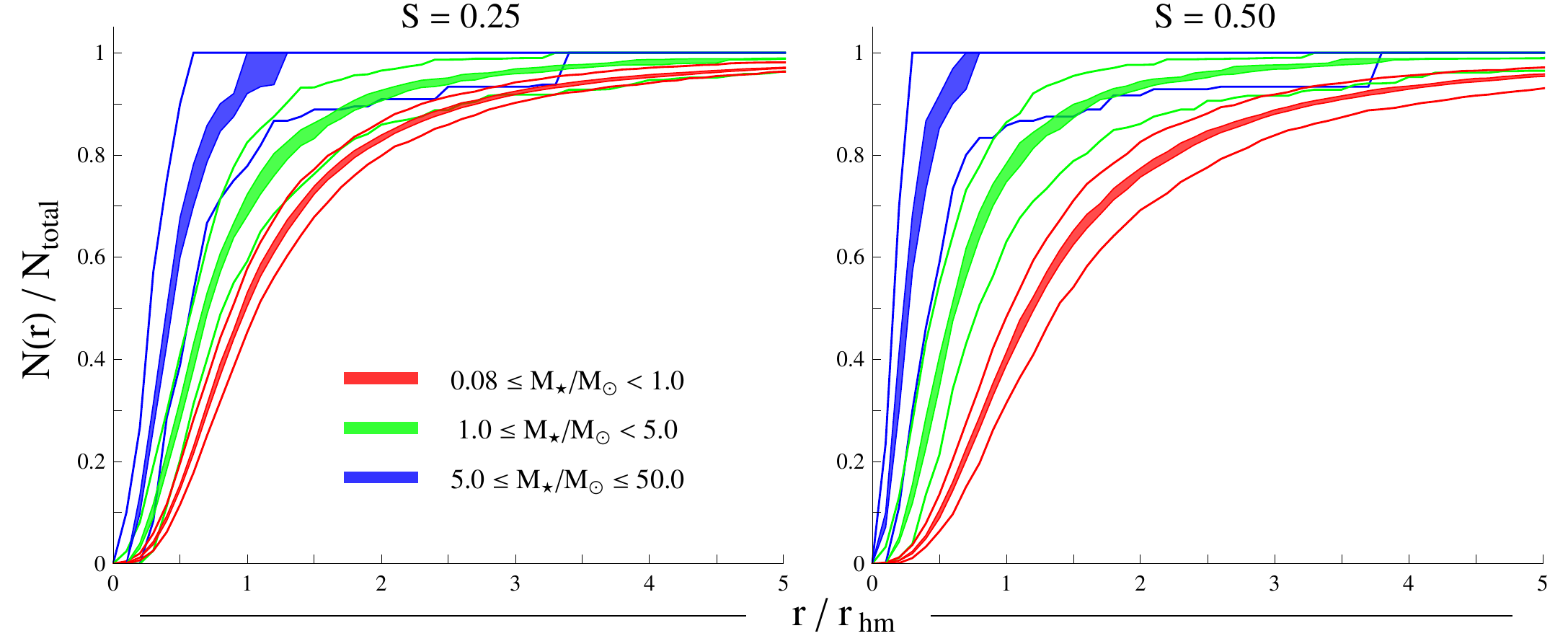}
 \caption{Mass segregation from initial conditions generated using the scheme of \citet{subr08}.  {\em Left}: moderate segregation, using their $S=0.25$.  {\em Right}: heavy segregation, using their $S = 0.5$.
 \label{S08Fig}}
\end{figure*}

A comparison of the cumulative distribution functions from the simulations starting from a uniform number density (Figure \ref{UniformResultsFig}) and from an isothermal density distribution (Figure \ref{IsothermalResultsFig}) shows a striking similarity between the two cases.  The only clear difference is in the shape of the initially unsegregated series, with the uniform number density case showing more final segregation than the isothermal mass density set.  This similarity, despite very different initial setups, argues for the robustness of the results to variations in the initial conditions.  

A source of potential concern is in our parametrization of mass segregation.  Our scheme sorts masses radially according to mass and then stretches the radii to match a chosen distribution, with no concern for the underlying stability or resultant dynamical structure.  One could argue that the star formation process itself may be likewise uncaring when it comes to the initial setup of a cluster; a young cluster is not expected to be in a relaxed state immediately following the expulsion of gas and its entrance into a stellar-dynamically dominated phase.  Nonetheless, a comparison to an alternative mass segregation scheme would seem warranted to ensure that our results are not dominantly a reflection of instabilities in our initial conditions.

\citet{subr08} (hereafter S08) developed a method for constructing initial conditions corresponding to dynamically evolved, segregated clusters.  We chose not to adopt their method for this work because we are concerned with very young clusters and primordial versus dynamical segregation.  Their system sets up the cluster in a stable, virialised state, and is parametrized by an index $S$.  With $S=0$ they have an unsegregated Plummer sphere, and $S=0.5$ yields a heavily segregated cluster with a central density that is approximately isothermal.  Using the code {\sc PLUMIX}\footnote{Authored by L. {\v S}ubr, and available at http://www.astro.uni-bonn.de.} we generated initial conditions corresponding to $S = 0.25$ and $S = 0.5$ in S08.  We then applied the same cumulative distribution analysis to these initial conditions as we used with our simulations.  These distributions are shown in Figure \ref{S08Fig}.

Their $S = 0.5$ has a near-isothermal mass distribution, so comparison to our series C2 would be the most equivalent.  The distributions from S08 bear little relation to our initial conditions, which is not unexpected due to the very different parametrization than used here.  However, after one crossing time the distributions appear remarkably similar.  Our initial conditions, constructed without regard to stability, seem to relax within one crossing time to the stable solutions of S08.  Further evolution confirms this, as seen in a comparison of the half-mass radii for series C2 in our Figure \ref{RhmIsoFig} with figure 5 of S08.  We therefore see that our initial conditions -- with very different initial density profiles showing the same trends in the evolution of mass segregation -- quickly (within one crossing time) approach a form corresponding to a dynamically mature state of segregation.  This level of segregation is maintained for several crossing times, as shown in our simulations and those in S08.  We conclude that strong primordial mass segregation, largely independent of its paramaterization, is unlikely to be dynamically erased over the first few Myr of cluster evolution.

\section{Model Limitations and simplifications}
In order to keep this study compact and tractable, we have made several simplifying assumptions.  A realistic star cluster will display some or all of the following complicating factors, which would affect these results to varying degrees.

\subsection{Primordial Binarity}
We have assumed that the initial binary fraction is zero, a situation at odds with observation and theory, which show that binarity is established to some extent during a stellar system's pre-main-sequence evolution \citep[e.g.][]{mathieu94,bate09}.  Close encounters between binaries and other clusters members can alter the two-body relaxation that moves the system toward equipartition (and thus mass segregation), and could perhaps change the rate of mixing in these simulations.  However, the short timescale and moderate stellar densities and velocities in these simulations suggest that the rate of close encounters will be low, and the effect of binaries will be minimal.  \citet{adams06} tested the non-importance of binarity in similarly sized clusters as ours, over a longer time period, and verified this argument.

\subsection{Gas Removal}
We begin our integrations from a gas-free state.  The details of gas removal will, in reality, affect a cluster to the point of determining its survival as a bound entity \citep{baumgardt07}.  The parameter space of gas removal details is quite large, and not tightly constrained.  In {\it n}-body simulations, it is typically modeled by globally lowering the gas mass, effectively removing its potential, over some timescale \citep[e.g.][]{adams06, baumgardt07, throop08}.  Generally, the effect of gas removal is to leave the stars moving super-virially, with consequent cluster expansion.  Global expansion of the clusters will have the qualitative effect of increasing the amount of mixing at large radii, as potentially fast-moving stars from the interior catch up with the expanding cluster outskirts, while the inner region retains whatever segregation it has as the interaction rate between stars drops.  This is seen to a mild extent in our series C runs, both of which display some level of global expansion, and which retain their segregation inwards of 2 $r_{hm}$.

\subsection{The Galactic Environment}
These experiments have taken place in a level of isolation that can only be found on a computer.  The importance, or non-importance, of the galactic gravitational field can be estimated by comparing the cluster size to the tidal radius $r_t$, which for a cluster of mass $M_{c}$ on a circular galactic orbit of radius $d$ we approximate as the Jacobi limit \citep{binney87},
\begin{equation}
  r_t = d \left ( \frac{M_{c}}{3 M_{g}(d)} \right )^{1/3},
\end{equation}
with $M_{g}(d)$ the galactic mass interior to the orbit.  Our clusters all have initial half-mass radii of 0.5 pc, with mass $M_c \sim 500$ \msun.  In a galaxy similar to the Milky Way, we then have the ratio $r_{hm} / r_t \le 0.06$ for galactic radii $d \geq 6$ kpc.  \citet{baumgardt07} showed that for $r_{hm}/r_t $ less than this value the cluster evolution proceeds as if it were isolated, so the effect of a galactic tide is unlikely to affect these results for star formation in the solar neighborhood.  Nearer the galactic center, tidal effects could be more important.  At $d=2$ kpc, for instance, $r_{hm} / r_t \sim 0.15$, making the assumption of isolation slightly suspect.  
Thus at later times in these simulations the exterior of a cluster near the galactic center may start to be lost to the field, but for the short timescales considered here the effect should be modest.

\subsection{Cluster Clumpiness}
Perhaps the most significant approximation made in this study is the spherical symmetry of the clusters.  While a spherical system offers appealing simplicity and a dramatic reduction of parameter space, star formation in nature tends to occur in a form that may better be approximated by a cylinder, a triaxial spheroid, or more complex filamentary or fractal geometries.  Breaking spherical symmetry (or introducing a bulk motion to parts of the cluster) makes analysis of quantities binned by radius problematic, as these may be an unsatisfactory average of different regions that have evolved to dynamically dissimilar ages.  

The build up of a cluster by merging sub-clusters is seen in some large-scale hydrodynamic simulations of cluster formation \citep{bonnell03}.  The question of mass segregation in such a process has seen preliminary exploration {\it n}-body exploration by \citet{mcmillan07}.  Interestingly, they show that the level of segregation present in the pre-merger clumps is inherited by the merger product, whether this segregation is primordial or dynamically realized on the shorter timescales associated with lower-{\it n} sub-clusters.  This yields a large, mass-segregated cluster that appears to be several dynamical times older than its `true' age.  The most relevant aspect of McMillan et al.'s result to this study is that the merging process, akin to violent relaxation, does not change the level of pre-existing mass segregation in a heirarchical system.

However, the simulations of McMillan et al.\ are not directly comparable to ours.  Firstly, their clusters contained over an order of magnitude more stars than the ones presented here.  More importantly, their initially segregated sub-clusters were generated by evolving an unsegregated cluster in isolation before merging, a process that treats primordial segregation as the {\em stable end point} of a dynamical process.  This work treats primordial segregation as an {\em imposed and unstable} initial condition, at a higher level of segregation than theirs.  

It would be worthwhile (though well outside the scope of this work) to explore the question of merging sub-clusters with the type of segregation we have used, to see if the violent relaxation process mixes the mass components more thoroughly than the two-body dynamics considered here.  If so, this process could ameliorate the inconsistency between observations and a strong mass-radius dependence in star formation mechanisms that we discuss below.  However, if  the segregation of the merged product remains at the dynamically unattainable levels imposed by our initial conditions, the nature of our results would be unchanged by including sub-cluster mergers.

\section{Comparison to observations}
Mass segregation is observed to some degree in many clusters; while there are clear examples like the potentially extreme cases or the Arches or 30 Doradus, here we are concerned with clusters in the earliest stages of their dynamical evolution.  Young clusters (less than a few Myr old) displaying some form of mass segregation include\footnote{Though see the cautionary note of \citet{ascenso08} regarding evidence for mass segregation.}: Mon R2 and NGC 2024 \citep{carpenter97,carpenter08}, which show little evidence for general segregation but seems to have a central concentration of the most massive sources; IC 1805 \citep{sagar88}; NGC 1893 \citep{sharma07}, showing segregation above $\sim 5$ \msuns which is argued to be primordial; NGC 6530 \citep{mcnamara86}; NGC 6231 \citep{raboud98a}; Stock 8 \citep{jose08}; and the ONC \citep{hillenbrand98}.

\begin{figure*}
 \includegraphics[width=140mm]{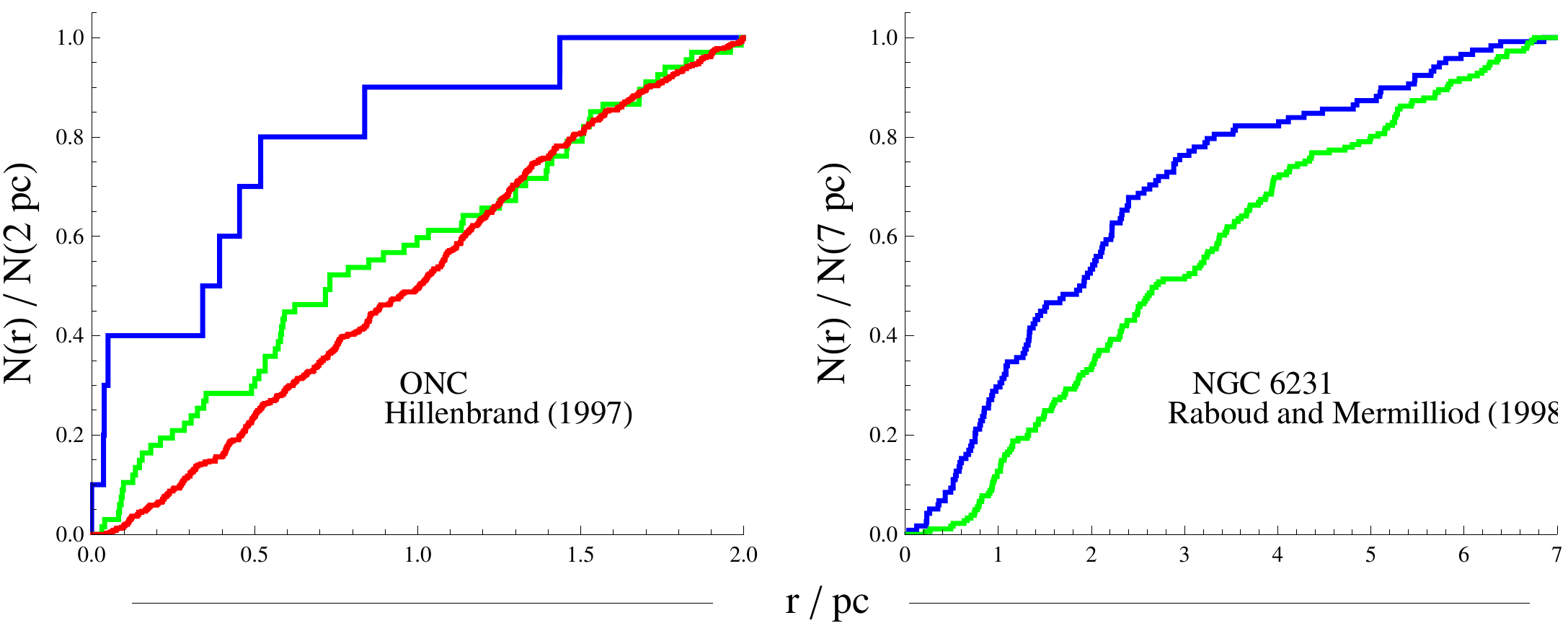}
 \caption{Cumulative fractional distributions for two young star clusters exhibiting mass segregation.  {\em Left}: the ONC, with data from \citet{hillenbrand97} and at an assumed distance of 400 pc.  Coloured lines correspond to the mass bins found throughout the text.  {\em Right}: NGC 6231, with data from \citet{raboud98a}.  Data is only available for stars more massive than 1 \msun, and we show only the moderate- and high-mass bins.\label{ObsCompFig}}
\end{figure*}

Here we compare observations of two clusters to the type of segregation seen in the simulations.  The clusters we consider are: the ONC, around 1 Myr old and the closest site of massive star formation at $\sim 400$ pc \citep[see the discussion of distance in][]{muench08}; and NGC 6231, a young (3-4 Myr) open cluster 1.8 kpc away \citep{raboud97}.
These clusters are chosen for their youth and the existence of published data sets that permit construction of cumulative fractional distributions for different mass components.  In Figure \ref{ObsCompFig} we show the cumulative distributions for these sources.  It should be mentioned that the radius quoted in the observations here is the {\em projected} radius, rather than the three-dimensional radius shown in our plots.  While we have chosen to retain the full spatial information in our simulations, there is very little change to the relative shape of the distributions.

\citet{hillenbrand98} discuss mass segregation in the ONC based on data from \citet{hillenbrand97}, which we use to construct cumulative distribution functions for the same mass components as in our simulations.  The ONC shows a clear segregation of stars more massive than 5 \msuns compared to the lower-mass stars, with apparent separation of the low- and moderate-mass bins inwards of around 1 pc.  Two-sample Kolmogorov--Smirnov (KS) tests of the distributions confirm that the high-mass distribution is probably distinct from the low- and moderate-mass components, with {\it p}-values of 0.003 and 0.027 respectively.  The separation between the moderate- and low-mass distributions is statistically unconvincing, with a {\it p}-value of 0.087.  

Mass segregation in NGC 6231 is considered by \citet{raboud98a}.  Data is only available for stars more massive than 1 \msun, so we show only the moderate- and high-mass distributions for this cluster.  The difference between the distributions of these two populations is confirmed with a KS test giving a {\it p}-value of $1.4\times 10^{-4}$.   Raboud et al.\ divide the stars for this data set into several mass bins and show evidence for increasing levels of segregation with increasing mass.  We note that if the incompleteness in the sample is a function of mass only, and is not radially dependent, than in principle the underlying distribution of stars as traced by the cumulative fraction should remain unchanged.  If this is the case for NGC 6231, then it is an example of a cluster with more general mass segregation at a young age.   

Looking only at the difference between high-mass stars and the rest of the stellar population, any of our models might be thought to show a similar degree of mass segregation at times greater than a few crossing times.  The best discriminant between our models is the separation between the low- and moderate-mass distribution.   This characteristic is not seen in the ONC data with any degree of certainty.  Series B and C, in both the uniform number density and isothermal mass density sets, show a clear distinction between the high-mass distribution and the other stars.  Series A, by contrast, shows very little difference between the low- and moderate-mass profiles.  Series D, beginning with only the massive stars segregated, likewise appears to be consistent with the available observations.

\section{Implications for Star Formation Theory}
The state of star formation theory (especially at the high end of the mass function) could not be described as quiescent, with two main theories currently at odds over the root mechanism of gathering stellar mass from the parent cloud, and how this mass is distributed amongst the stars to create the initial mass function.  The competitive accretion theory \citep{bonnell01a} posits that fragmentation occurs at a characteristic mass set by the Jeans instability, and accretion onto these protostellar seeds leads naturally to a the spread in masses, with massive stars being those that were born in advantageous locations.  The core collapse model \citep{mckee03} argues for a direct mapping between the mass of a collapsing core and the mass of the star that emerges-- the mass is assembled prior to collapse, not during the early dynamics of an embedded cluster.  Massive cores are supported against collapse by turbulence and the high pressure in massive star forming regions.  

Taking both of these theories at face value and at their simplest conceptual limit, one would argue for primordial mass segregation.  In competitive accretion in a spherical system, the richest gas reservoir is deep in the cluster potential.  Protostars moving slowly through the cluster center would accrete more material than on the outskirts, leading to an initially segregated cluster.  The core collapse model leads to the same situation.  The fiducial high-mass core presented in \citet{mckee03} supposes that the high mass core is located well inside the half-mass radius of the parent clump, and it is loosely assumed that the core mass traces the clump mass to some degree.  If stars acquire their mass from a core and don't move far during formation, and the core mass is tied to the clump density profile, a degree of continuous mass segregation follows naturally.
In an idealized spherical cluster, both theories would thus seem to argue for a cluster initially well-segregated, with massive stars preferentially forming near the cluster center.

These results would argue that such a low-level view of the theories is unlikely to be reflective of an actual mode of star formation.  Observations of the youngest clusters, combined with the level of mixing seen in our simulations, do not support a simple spherical cluster with a purely radial dependence on the final mass of a star.  Understanding star formation on a cluster-scale appears to require levels of substructure not reflected in a spherical theory, and perhaps only revealed through cluster-scale simulations.  For example, the simulation of a non-spherical GMC collapse in \citet{bonnell08} displays sub-clustering in the stellar positions, an arrangement not considered in this work.

Despite the lack of segregation across all mass ranges, it is clear that the most massive stars tend to be centrally concentrated at an early age.  The Trapezium stars are an obvious example, as well as Mon R2, NGC 2024 and NGC 1893.  In Stock 8, cumulative distributions of massive and low-mass stars are statistically different if the mass cut is 4 \msun, while they are indistinguishable if the mass cut is 1 \msun, suggesting that the most massive stars are segregated while the moderate and low-mass stars are more evenly distributed \citep{jose08}.  \citet{raboud98a} argue for a similar state for NGC 6231, though some segregation of the moderate-mass stars is evident.  \citet{allison09} apply a different method, minimum spanning trees, to the ONC and find no evidence for segregation below 5 \msun.
A promising route to this arrangement is formation in small subclusters that either form segregated or quickly dynamically segregate, and merge leaving behind a subcluster of massive stars in the `halo' of a less-segregated cluster.  This process has been seen in cluster-scale simulations and has seen preliminary {\it n}-body exploration by \citet{mcmillan07}, though more thorough experiments to test the degree of post-merger segregation have not yet been performed. 

The alternative to merging subclusters is a fundamentally different mode of formation for high-mass stars, or special conditions found only at the centers of clusters that lead to high-mass formation.  This would remove the continuous segregation across all masses and leave only the most massive stars initially segregated.  Observations of young, embedded massive groupings like W3 IRS 5 \citep{megeath05} might suggest that the tight clustering of massive stars is indeed a primordial effect.  The results of our series D, testing a qualitatively similar scenario, are consistent with observations of older (but still young) clusters. 

In summary, dynamical mixing in young clusters is insufficient to erase the signature of a continuous primordial mass segregation that affects stars of all masses.  Initially unsegregated initial conditions evolve to a state that broadly matches the level of segregation observed, but as shown previously \citep{bonnell98} can not account for the presence of dense subclusters like the Trapezium.  A star-formation scenario in which only the most massive stars are primordially segregated is consistent with observations, and offers a way to account for compact groups of young, massive stars.

\section*{Acknowledgments}
We acknowledge STFC for support of this work.
NM gratefully thanks the midwives, nurses and doctors at the Ninewells Hospital NICU for caring for his son, Calvin Ajax, born during the revision of this paper.

\bibliographystyle{mn2e}

\begin{thebibliography}{}

\bibitem[\protect\citeauthoryear{{Aarseth}}{{Aarseth}}{2003}]{aarseth03}
{Aarseth} S.~J.,  2003, {Gravitational N-Body Simulations}.
Gravitational N-Body Simulations, by Sverre J.~Aarseth, pp.~430.~ISBN
  0521432723.~Cambridge, UK: Cambridge University Press, November 2003.

\bibitem[\protect\citeauthoryear{{Adams}, {Proszkow}, {Fatuzzo} \&
  {Myers}}{{Adams} et~al.}{2006}]{adams06}
{Adams} F.~C.,  {Proszkow} E.~M.,  {Fatuzzo} M.,    {Myers} P.~C.,  2006, \apj,
  641, 504

\bibitem[\protect\citeauthoryear{{Allison}, {Goodwin}, {Parker}, {Portegies
  Zwart}, {de Grijs} \& {Kouwenhoven}}{{Allison} et~al.}{2009}]{allison09}
{Allison} R.~J.,  {Goodwin} S.~P.,  {Parker} R.~J.,  {Portegies Zwart} S.~F.,
  {de Grijs} R.,    {Kouwenhoven} M.~B.~N.,  2009, arXiv:0901.2047

\bibitem[\protect\citeauthoryear{{Ascenso}, {Alves} \& {Lago}}{{Ascenso}
  et~al.}{2008}]{ascenso08}
{Ascenso} J.,  {Alves} J.,    {Lago} M.~T.~V.~T.,  2008, ArXiv e-prints

\bibitem[\protect\citeauthoryear{{Bate}}{{Bate}}{2009}]{bate09}
{Bate} M.~R.,  2009, \mnras, 392, 590

\bibitem[\protect\citeauthoryear{{Baumgardt} \& {Kroupa}}{{Baumgardt} \&
  {Kroupa}}{2007}]{baumgardt07}
{Baumgardt} H.,  {Kroupa} P.,  2007, \mnras, 380, 1589

\bibitem[\protect\citeauthoryear{{Binney} \& {Tremaine}}{{Binney} \&
  {Tremaine}}{1987}]{binney87}
{Binney} J.,  {Tremaine} S.,  1987, {Galactic dynamics}.
Princeton, NJ, Princeton University Press, 1987, 747 p.

\bibitem[\protect\citeauthoryear{{Bonnell}, {Bate}, {Clarke} \&
  {Pringle}}{{Bonnell} et~al.}{2001}]{bonnell01a}
{Bonnell} I.~A.,  {Bate} M.~R.,  {Clarke} C.~J.,    {Pringle} J.~E.,  2001,
  \mnras, 323, 785

\bibitem[\protect\citeauthoryear{{Bonnell}, {Bate} \& {Vine}}{{Bonnell}
  et~al.}{2003}]{bonnell03}
{Bonnell} I.~A.,  {Bate} M.~R.,    {Vine} S.~G.,  2003, \mnras, 343, 413

\bibitem[\protect\citeauthoryear{{Bonnell}, {Clark} \& {Bate}}{{Bonnell}
  et~al.}{2008}]{bonnell08}
{Bonnell} I.~A.,  {Clark} P.,    {Bate} M.~R.,  2008, \mnras, 389, 1556

\bibitem[\protect\citeauthoryear{{Bonnell} \& {Davies}}{{Bonnell} \&
  {Davies}}{1998}]{bonnell98}
{Bonnell} I.~A.,  {Davies} M.~B.,  1998, \mnras, 295, 691

\bibitem[\protect\citeauthoryear{{Carpenter} \& {Hodapp}}{{Carpenter} \&
  {Hodapp}}{2008}]{carpenter08}
{Carpenter} J.,  {Hodapp} K.,  2008, ArXiv e-prints

\bibitem[\protect\citeauthoryear{{Carpenter}, {Meyer}, {Dougados}, {Strom} \&
  {Hillenbrand}}{{Carpenter} et~al.}{1997}]{carpenter97}
{Carpenter} J.~M.,  {Meyer} M.~R.,  {Dougados} C.,  {Strom} S.~E.,
  {Hillenbrand} L.~A.,  1997, \aj, 114, 198

\bibitem[\protect\citeauthoryear{{Hillenbrand}}{{Hillenbrand}}{1997}]{hillenbr%
and97}
{Hillenbrand} L.~A.,  1997, \aj, 113, 1733

\bibitem[\protect\citeauthoryear{{Hillenbrand} \& {Hartmann}}{{Hillenbrand} \&
  {Hartmann}}{1998}]{hillenbrand98}
{Hillenbrand} L.~A.,  {Hartmann} L.~W.,  1998, \apj, 492, 540

\bibitem[\protect\citeauthoryear{{Jose}, {Pandey}, {Ojha}, {Ogura}, {Chen},
  {Bhatt}, {Ghosh}, {Mito}, {Maheswar} \& {Sharma}}{{Jose}
  et~al.}{2008}]{jose08}
{Jose} J.,  {Pandey} A.~K.,  {Ojha} D.~K.,  {Ogura} K.,  {Chen} W.~P.,  {Bhatt}
  B.~C.,  {Ghosh} S.~K.,  {Mito} H.,  {Maheswar} G.,    {Sharma} S.,  2008,
  \mnras, 384, 1675

\bibitem[\protect\citeauthoryear{{Khalisi}, {Amaro-Seoane} \&
  {Spurzem}}{{Khalisi} et~al.}{2007}]{khalisi07}
{Khalisi} E.,  {Amaro-Seoane} P.,    {Spurzem} R.,  2007, \mnras, 374, 703

\bibitem[\protect\citeauthoryear{{Kim}, {Figer}, {Kudritzki} \&
  {Najarro}}{{Kim} et~al.}{2006}]{kim06}
{Kim} S.~S.,  {Figer} D.~F.,  {Kudritzki} R.~P.,    {Najarro} F.,  2006, \apjl,
  653, L113

\bibitem[\protect\citeauthoryear{{Kroupa}}{{Kroupa}}{2001}]{kroupa01}
{Kroupa} P.,  2001, \mnras, 322, 231

\bibitem[\protect\citeauthoryear{{Mathieu}}{{Mathieu}}{1994}]{mathieu94}
{Mathieu} R.~D.,  1994, \araa, 32, 465

\bibitem[\protect\citeauthoryear{{McKee} \& {Tan}}{{McKee} \&
  {Tan}}{2003}]{mckee03}
{McKee} C.~F.,  {Tan} J.~C.,  2003, \apj, 585, 850

\bibitem[\protect\citeauthoryear{{McMillan}, {Vesperini} \& {Portegies
  Zwart}}{{McMillan} et~al.}{2007}]{mcmillan07}
{McMillan} S.~L.~W.,  {Vesperini} E.,    {Portegies Zwart} S.~F.,  2007, \apjl,
  655, L45

\bibitem[\protect\citeauthoryear{{McNamara} \& {Sekiguchi}}{{McNamara} \&
  {Sekiguchi}}{1986}]{mcnamara86}
{McNamara} B.~J.,  {Sekiguchi} K.,  1986, \apj, 310, 613

\bibitem[\protect\citeauthoryear{{Megeath}, {Wilson} \& {Corbin}}{{Megeath}
  et~al.}{2005}]{megeath05}
{Megeath} S.~T.,  {Wilson} T.~L.,    {Corbin} M.~R.,  2005, \apjl, 622, L141

\bibitem[\protect\citeauthoryear{{Muench}, {Getman}, {Hillenbrand} \&
  {Preibisch}}{{Muench} et~al.}{2008}]{muench08}
{Muench} A.,  {Getman} K.,  {Hillenbrand} L.,    {Preibisch} T.,  2008, ArXiv
  e-prints

\bibitem[\protect\citeauthoryear{{Portegies Zwart}, {Gaburov}, {Chen} \&
  {G{\"u}rkan}}{{Portegies Zwart} et~al.}{2007}]{portegies-zwart07}
{Portegies Zwart} S.,  {Gaburov} E.,  {Chen} H.-C.,    {G{\"u}rkan} M.~A.,
  2007, \mnras, 378, L29

\bibitem[\protect\citeauthoryear{{Raboud}, {Cramer} \& {Bernasconi}}{{Raboud}
  et~al.}{1997}]{raboud97}
{Raboud} D.,  {Cramer} N.,    {Bernasconi} P.~A.,  1997, \aap, 325, 167

\bibitem[\protect\citeauthoryear{{Raboud} \& {Mermilliod}}{{Raboud} \&
  {Mermilliod}}{1998}]{raboud98a}
{Raboud} D.,  {Mermilliod} J.-C.,  1998, \aap, 333, 897

\bibitem[\protect\citeauthoryear{{Sagar}, {Miakutin}, {Piskunov} \&
  {Dluzhnevskaia}}{{Sagar} et~al.}{1988}]{sagar88}
{Sagar} R.,  {Miakutin} V.~I.,  {Piskunov} A.~E.,    {Dluzhnevskaia} O.~B.,
  1988, \mnras, 234, 831

\bibitem[\protect\citeauthoryear{{Sharma}, {Pandey}, {Ojha}, {Chen}, {Ghosh},
  {Bhatt}, {Maheswar} \& {Sagar}}{{Sharma} et~al.}{2007}]{sharma07}
{Sharma} S.,  {Pandey} A.~K.,  {Ojha} D.~K.,  {Chen} W.~P.,  {Ghosh} S.~K.,
  {Bhatt} B.~C.,  {Maheswar} G.,    {Sagar} R.,  2007, \mnras, 380, 1141

\bibitem[\protect\citeauthoryear{{Spitzer} Jr. \& {Shull}}{{Spitzer} \&
  {Shull}}{1975}]{spitzer75}
{Spitzer} Jr. L.,  {Shull} J.~M.,  1975, \apj, 201, 773

\bibitem[\protect\citeauthoryear{{Spitzer}}{{Spitzer}}{1969}]{spitzer69}
{Spitzer} L.~J.,  1969, \apjl, 158, L139

\bibitem[\protect\citeauthoryear{{Throop} \& {Bally}}{{Throop} \&
  {Bally}}{2008}]{throop08}
{Throop} H.~B.,  {Bally} J.,  2008, \aj, 135, 2380

\bibitem[\protect\citeauthoryear{{{\v S}ubr}, {Kroupa} \& {Baumgardt}}{{{\v
  S}ubr} et~al.}{2008}]{subr08}
{{\v S}ubr} L.,  {Kroupa} P.,    {Baumgardt} H.,  2008, \mnras, 385, 1673

\end{thebibliography}

\bsp

\label{lastpage}

\end{document}